%% file: main.tex
\pdfoutput=1
\PassOptionsToPackage{hidelinks}{hyperref}
\documentclass[letterpaper,twocolumn,10pt]{article}

\usepackage{usenix,epsfig,endnotes}
\usepackage{graphicx}
\usepackage{color}
\usepackage{hyperref}
\usepackage{listings}
\usepackage{xspace}
\usepackage{colortbl}
\usepackage{siunitx}

\usepackage{float}

\usepackage{amsmath}
\usepackage{amsfonts}
\usepackage{amsthm}
\usepackage{enumitem}
\usepackage{pythonhighlight}

\newfloat{lstfloat}{t!}{lop}
\floatname{lstfloat}{Listing}

\newcommand{\brainslug}{\textsc{BrainSlug}\xspace}

\begin{document}
\date{}
\title{BrainSlug: Transparent Acceleration of Deep Learning Through Depth-First Parallelism}
\author{
	{\rm Nicolas Weber,}
	{\rm Florian Schmidt,}
	{\rm Mathias Niepert,}
	{\rm Felipe Huici}
	\\NEC Laboratories Europe, Systems and Machine Learning Group
}

\maketitle

\input{abstract.tex}
\input{introduction.tex}
\input{background.tex}
\input{method.tex}
\input{implementation.tex}
\input{eval.tex}
\input{related.tex}
\input{discussion.tex}

\input{references.tex}

\end{document}

%% file: abstract.tex
\section*{Abstract}
Neural network frameworks such as PyTorch and TensorFlow are the
workhorses of numerous machine learning applications ranging from
object recognition to machine translation. While these frameworks are
versatile and straightforward to use, the training of and inference in
deep neural networks is resource (energy, compute, and memory) intensive.

In contrast to recent works focusing on algorithmic enhancements, we
introduce \brainslug, a framework that \emph{transparently}
accelerates neural network workloads by changing the default
layer-by-layer processing to a depth-first approach, reducing the
amount of data required by the computations and thus improving the
performance of the available hardware caches.
\brainslug achieves performance improvements of up
to 41.1\% on CPUs and 35.7\% on GPUs. These optimizations come at zero cost
to the user as they do not require hardware changes and only need tiny
adjustments to the software.

%% file: introduction.tex
\section{Introduction}
\label{sec:intro}
Artificial intelligence, and neural networks in particular, have
gained immense notoriety in the past few years. Their flexibility
means that they can be applied to a wide range of applications, from
recommendation systems for online stores, to autonomous driving,
to financial fraud detection or optimization of production lines. 

One of the main issues with neural networks is their high
computational time. While over the years many algorithmic and
hardware optimizations to reduce the cost of these computations have
arisen, the sequential way in which a neural network is
processed, taking input data and computing on it from layer to layer
before moving onto the next input data, has remained largely
fixed. Even though this breadth-first processing is straightforward,
the fact that each pass through the network acts on a relatively large
amount of data causes constant cache trashing (whether on CPUs, GPUs,
or other hardware), reducing their effectiveness and ultimately
increasing computation time. This partly explains why processors with
extremely high memory throughput are used for neural networks so that
the processors are never idle.

In this report we propose the use of a depth-first approach: we take a
subset of the input data (e.g., a part of an image) that can fit in
L1 cache and compute all layers, then repeat the process for the next
subset of the data. At this high level the process sounds simple;
however, there are two main issues. First, only certain operations
(i.e., layer types) are able to function when given only a subset of
the data. Second, processing data in this way requires the user to
write specialized compute kernels for each possible sequence of
layers. This is clearly difficult to do by hand and points to the need
of an automated system to carry this out.

We implemented \brainslug (\href{http://brainslug.info}{brainslug.info}), a system that enables depth-first
computation of neural networks, providing \emph{transparent}
acceleration through improvements to data locality. We make the
following specific contributions:

\begin{itemize}[noitemsep,nolistsep]
  \item A novel, depth-first method for neural network computation
    that increases data locality and reduces computation time. The
    method does not change the actual results of the computation, and
    is widely applicable to a large set of neural networks and
    different types of hardware.

  \item An implementation of this method, including a modular
    architecture that allows for easy extensibility to multiple neural
    network frameworks (e.g., PyTorch~\cite{pytorch}, Theano~\cite{theano}, Caffe~\cite{caffe}, TensorFlow~\cite{tensorflow}) and hardware
    targets (e.g., CPUs, GPUs, FPGAs, etc.).

  \item The implementation and evaluation of \brainslug using a
    PyTorch front-end and CPU and GPU back-ends. Through extensive
    experimentation we show that \brainslug achieves speed-ups of up
    to 41.1\% on CPUs and 35.7\% on GPUs while requiring only tiny adjustments
    to a user's program.

\end{itemize}

In the following, we first give a brief background about neural
networks in Section~\ref{sec:background}. Section~\ref{sec:method}
describes the main idea behind \brainslug, followed in
Section~\ref{sec:implementation} by a description of the system's
implementation, and an in-depth evaluation in Section~\ref{sec:eval}.
Finally, we discuss related work in Section~\ref{sec:related} and
conclude in Section~\ref{sec:discussion} with a discussion and future work.

%% file: background.tex
\begin{figure}[t]
	\centering
	\includegraphics[width=\columnwidth]{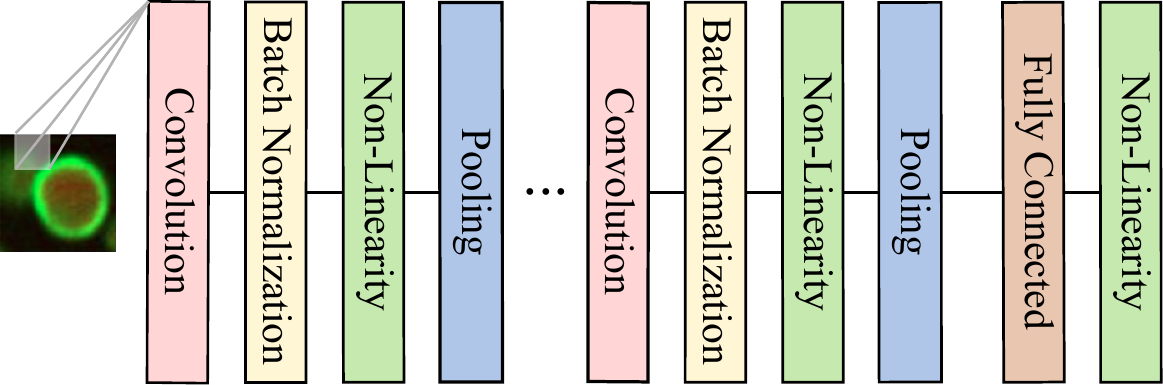}
	\caption{\label{fig-cnn} Typical instance of a deep convolutional
		neural network. All CNNs have convolutional and non-linearity
		layers. While the normalization and pooling layers are optional, they
		are found in almost all of the existing CNNs for computer vision
		tasks.}
\end{figure}

\section{Background on Neural Networks}
\label{sec:background}
On a basic level, a neural network corresponds to a sequence of
operations which act on numerical input data of a certain predefined size,
so-called tensors.  In computer vision, for instance, the input
tensors are typically three-dimensional data structures, with two
dimensions defining the two-dimensional picture (of $w \times h$
pixels), and the third dimension $d$ containing the information of
each color channel.  The naming of those three dimensions as width,
height, and channels has been adopted as a common general naming
scheme beyond the field of computer vision.

\begin{figure}[t]
	\centering
	\includegraphics[width=0.7\columnwidth]{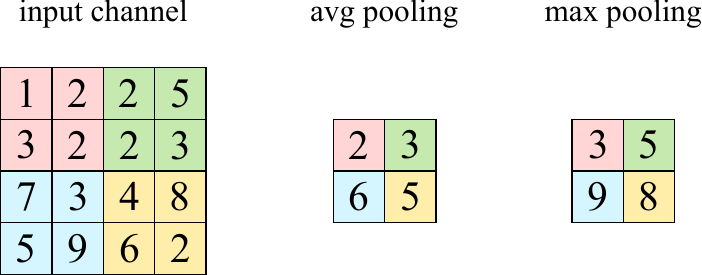}
	\vspace{1em}
	\caption{\label{fig-pool} Example of a pooling operation on
		one input channel. The aggregation is performed over
		non-overlapping regions of the input space. The figure is
		based on an example from the Caffe
		tutorial~\cite{jia2014caffe}.}
\end{figure}

\begin{figure}[t]
	\centering
	\includegraphics[width=0.75\columnwidth]{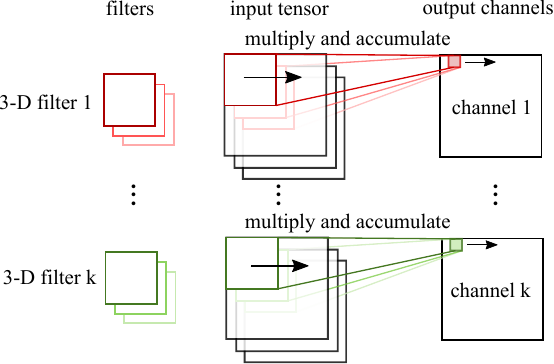}
	\vspace{1em}
	\caption{\label{fig-conv} Each convolutional layer moves a set
          of 2-D filters (each being part of one 3-D filter) over the
          input channels from left to right and top to bottom and
          multiplies and accumulates the corresponding values. }
\end{figure}

The transformations applied to the input data are grouped into layers,
with each layer executing a certain type of operation (an example of a
deep neural network is given in \autoref{fig-cnn}). The most common
layers are:\\
\noindent(1)~\textit{Element-wise layers} apply a function to each of the input
tensor's values independently. Typical examples of element-wise
layers are normalization and non-linearity layers. The former
normalize the output of a previous layer to conform to a specific
desired distribution, improving convergence behavior and accuracy of
the network. A \textit{non-linear or activation} layer applies an
activation function element-wise to the input. A commonly
used activation function is the rectified linear unit (ReLU), computing $f(x) = \max(0, x)$ on each input~\cite{nair2010rectified}.\\
(2)~\textit{Pooling} layers operate on predefined and fixed regions of
the input tensor. These predefined regions are often non-overlapping and square-shaped.
An example of two pooling operations (average and maximum) is given in
Figure~\ref{fig-pool}.\\
(3)~\textit{Convolutional} layers apply a convolution
operation.  In the image example, a convolutional
layer comprises $k$ groups of $d$ filters each, where in each group
one of the filters gets applied to each channel.  Since each filter
works on a 2-D channel, each of the $k$ groups is also called a 3-D
filter (\autoref{fig-conv}). The output dimension of a convolutional layer has depth $k$, while keeping the
channel dimensions $w \times h$.\\
(4)~\textit{Fully-connected} or \textit{dense} layers: Each value of
the output vector is a weighted sum of all input values, that is, all
output values are connected to all input values. In deep CNNs, there
are usually only a few dense layers and these
are located at the end of the network (see \autoref{fig-cnn}).

Each neural network can be mapped to a Directed Acyclic Graph (DAG)
whose nodes correspond to (groups of) element-wise operations and
whose edges represent the input-output relationships between the
computations.  Most deep learning frameworks map a given neural
network specification to such a computation graph and execute the
graph on one or multiple devices. Since the DAG has a unique root
(corresponding to the input data), every node has a unique depth. A
level of a computation DAG is made up of all operations at a specific depth. All
nodes at the same level compute their operations independently of each
other, but nodes at deeper levels might depend on results of nodes in
previous levels.  The computation DAG, therefore, also represents the
dependency structure of the computations. Figure~\ref{fig:trio-bf} illustrates the connection
between the layers of a neural network (right), the computation graph
(left), and the corresponding code snippet (center).

%% file: method.tex
\begin{figure*}[t]%
\centering
	\begin{minipage}[b]{0.35\textwidth}%
		\includegraphics[width=\columnwidth]{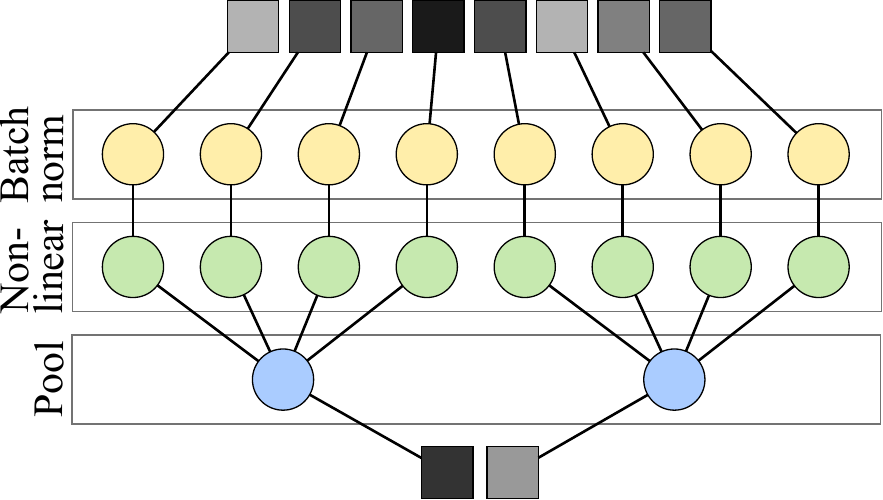}%
	\end{minipage}%
	\hspace{5mm}%
	\begin{minipage}[b]{0.36\textwidth}%
	\begin{lstlisting}[frame=none, numbers=none]
	
	foreach x in input:
		a = batchNorm(x)
		
	foreach x in a:
		b = max(x, 0)	
	
	output[0] = max(b[:4])
	output[1] = max(b[4:])
    \end{lstlisting}%
    \end{minipage}%
    \begin{minipage}[b]{0.25\textwidth}%
		\includegraphics[width=\columnwidth]{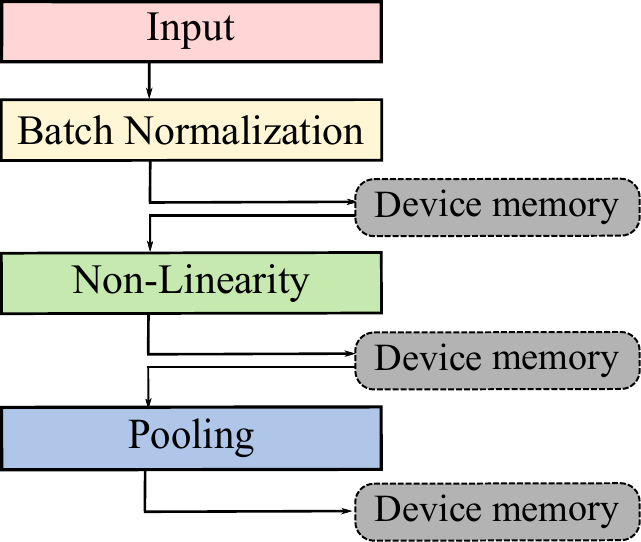}%
	\end{minipage}%
	\vspace{1em}
	\caption{\label{fig:trio-bf} Breadth-first parallelism. The computation graph (left) for the neural network layers (right) and the corresponding code snippet (middle). The standard method for processing the layers is in a breath-first manner, with every operation of level $i$ in the computation graph executed in parallel before operations at level $i+1$. This is indicated by the white boxes (left) surrounding each level of the computation graph.  }%

	\begin{minipage}[b]{0.35\textwidth}%
		\includegraphics[width=\columnwidth]{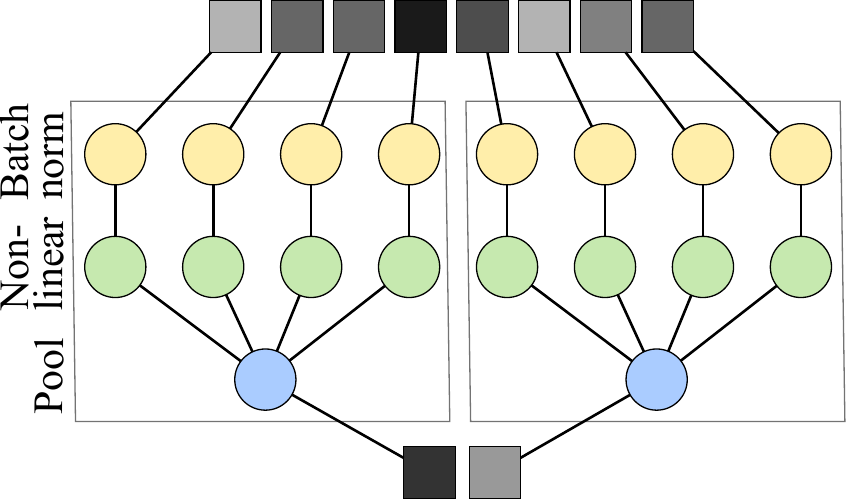}%
	\end{minipage}%
	\hspace{5mm}%
	\begin{minipage}[b]{0.36\textwidth}%
	\begin{lstlisting}[numbers=none, frame=none]
	foreach x in input[:4]:
		a[x] = batchNorm(x)
		b[x] = max(a[x], 0)	
	output[0] = max(b)
		
	foreach x in input[4:]:
		a[x] = batchNorm(x)
		b[x] = max(a[x], 0)	
	output[1] = max(b)
    \end{lstlisting}%
    \end{minipage}%
    \begin{minipage}[b]{0.26\textwidth}%
	\includegraphics[width=\columnwidth]{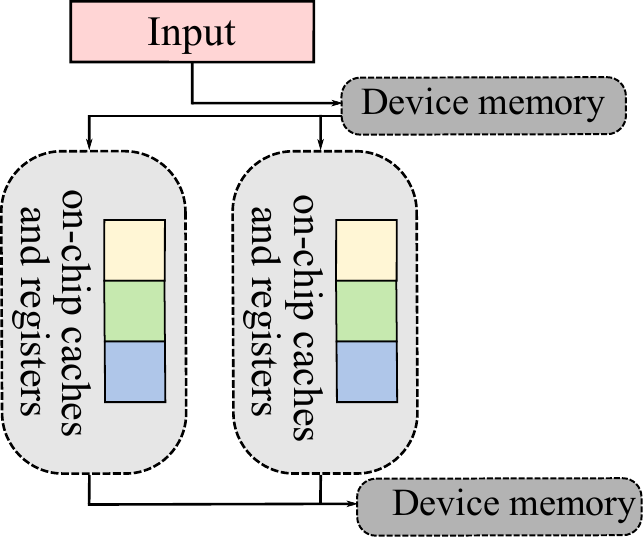}%
	\end{minipage}%
	\vspace{1.0em}%
	\caption{\label{fig:trio-df} Depth-first
				parallelism. The computation graph (left) for the
				neural network layers (right) and the corresponding code
				snippet (middle). \brainslug can detect independent
				paths in the computation graphs and aggregate these
				paths into independent processing blocks. The white
				boxes (left) indicate the parts of the computation graph
				that \brainslug chose to parallelize. The
				intermediate data generated within these blocks fits
				into the hardware cache.}%
\end{figure*}

\section{\textsc{BrainSlug}: Method Principles}
\label{sec:method}

Every DNN has to perform numerous passes through the network during
training and prediction. In many cases, millions of floating-point
operations are required for one pass and there are millions of passes
per training task. With \brainslug we want to improve the
resource utilization of DNN frameworks, with a focus on accelerating
(groups of) element-wise and pooling layers, all the while ensuring
that this acceleration is transparent to users and that it can be
implemented irrespective of the deep learning framework used (e.g.,
PyTorch, Theano, Caffe) and irrespective of the target hardware.

Towards this goal, we address the shortcoming of existing deep learning
frameworks that always execute neural networks layer by layer. The
dependency structure of the computation graph, however, allows us to
exploit situations in which a different execution pattern is possible.
Our work on \brainslug is therefore motivated by two basic questions:

\begin{enumerate}[noitemsep]
\item Are there ways to rearrange the standard execution order of the
  computation graph such that the hardware can operate more
  efficiently while delivering the same numerical results?

\item Is there a method for detecting when such a rearranging is
  possible such that it is both frequently applicable and efficient to
  compute?
\end{enumerate}

With the proposed \brainslug approach, we answer both questions in the
affirmative. First, we show that the way in which the computation
graph is executed has an impact on the efficiency of the
computations. In several situations, the same set of operations
can be executed such that the intermediate data fits into the
caches and registers of a device, circumventing the need to read and
write from the device's main memory.  This is possible by executing
\emph{independent paths} (or groups of independent paths) in the
computation graph in parallel, essentially parallelizing the computation
graph not only in a breadth-first but, when beneficial, in a
depth-first manner. Second, we show that for a large class of DNNs,
there is a generic method for detecting independent paths whose
intermediate data fits into the caches and registers.  

\begin{figure*}[t]%
\centering
	\begin{minipage}[b]{0.5\textwidth}%
		\includegraphics[width=0.85\linewidth]{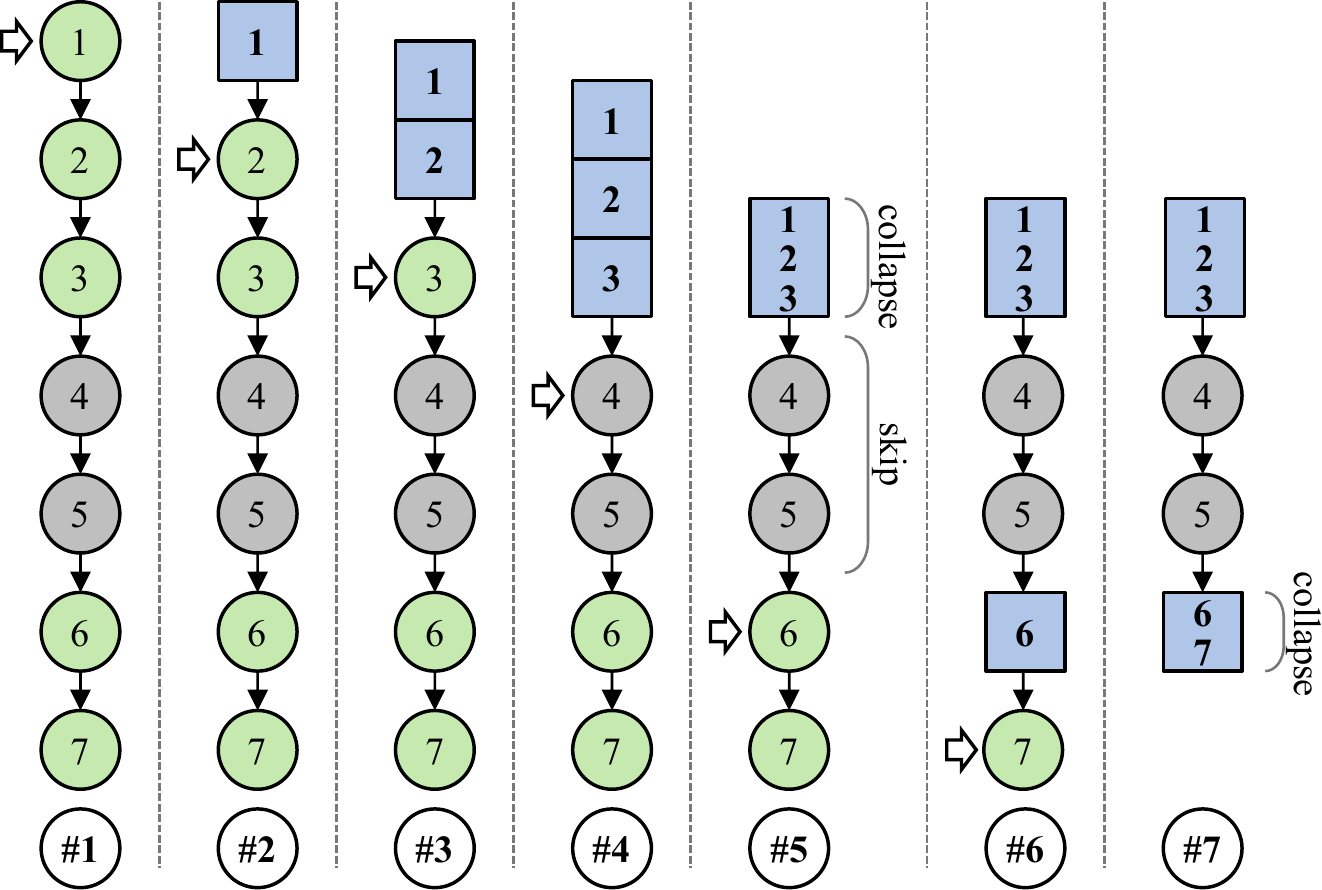}%
	\end{minipage}%
	\hspace{0.02\linewidth}%
	\begin{minipage}[b]{0.42\linewidth}%
		\begin{lstlisting}[frame=none]
stack = []
foreach(node in network):
	if(node is optimizable):
		if(stack is empty):
			network.replace(node, stack)
		else
			network.remove(node)
		stack.add(node)
	else:
		stack.collapse()
		stack = []
	\end{lstlisting}%
	\end{minipage}%
	\vspace{1em}%
	\caption{\brainslug's layer aggregation process, shown visually (left) and as pseudo-code (right).}%
	\label{fig:network-parsing}%
\end{figure*}

\subsection{Depth-First Parallelism}
Existing deep learning frameworks parallelize the computation graph in a
breadth-first manner, finishing all computations at one level of computation
before starting the next level's computations. We refer to this as
\emph{breadth-first parallelism} because the operations at one DAG level are
executed in parallel before the computation proceeds to lower DAG levels.  

\brainslug is able to detect and execute more complex independent
computation paths in parallel. While this does not reduce the overall
number of operations, it often leads to situations where the data
accessed by these independent paths fit into the caches and
registers of the hardware, increasing performance.  We refer to the
parallel processing of independent paths in the computation graph as
\emph{depth-first parallelism}. The challenge is now to transparently
detect and interleave these parallelism types to suit the
characteristics of particular hardware.

\autoref{fig:trio-df} illustrates depth-first parallelism for the
example in \autoref{fig:trio-bf}. The computation graph is the same but the operations
are grouped according to the independent computation paths involving
the normalization and non-linear operations, which are merged in the
pooling layer. Whereas in the typical breath-first, layer-wise
execution of DNNs the data has to be written and read from main memory
for each layer, here the intermediate data are small enough to fit
into the hardware's cache, improving overall performance.
Fortunately, as we show in what follows, detecting the existence of
these parallel computation paths in the DAGs of DNNs is both efficient
and frequently possible.

\subsection{Aggregation Detection}
To detect layers that can be aggregated, \brainslug parses through the DAG of a given neural
network layer-by-layer and identifies sequences of layers that support data
\emph{locality}, that is, sequences of layers that operate on a sub-set of the data (e.g., a
portion of an image) reducing the number of input-output dependencies in the computation graph. Examples of such layers
are layers performing element-wise and pooling operations.  We add all
consecutive sequences of such layers to a \emph{stack} (see \autoref{fig:network-parsing}),
which we then \emph{collapse} by analyzing the underlying computations and rewriting
them to utilize depth-first parallelism. A stack, therefore, partitions
independent computation paths in the DAG into paralellizable code blocks such
that each such block's intermediate data fits into the device caches. As each of the SIMD units of a single
device core share the same cached data,
we need to find an input data region (and the corresponding combined paths in the computation DAG)  that (1) is big enough to
keep all SIMD units utilized during the computations and (2) does not exceed the
cache size limit. As one can see in \autoref{fig:trio-df}, if one SIMD
unit is mapped onto one white box, we need at least the same number of
boxes as we have SIMD units to operate efficiently.

During the rewriting of the computations we need to take the
dependencies of the original DAG into account. For example, as can be
seen in \autoref{fig:trio-df}, the pooling layer requires all data
from the element-wise layers to be calculated before it can
perform its computations. 

 We provide a
detailed description of all of \brainslug's mechanisms in the
following sections.

%% file: implementation.tex
\section{\brainslug: Architecture and Implementation}
\label{sec:implementation}
One of the explicit goals of \brainslug is to transparently accelerate
neural networks (NN) \emph{irrespective} of the framework (e.g., PyTorch, Theano,
Caffe) they are implemented in. Further, we want the acceleration to 
apply to a wide range of hardware devices including GPUs, CPUs, FPGAs, and vector
processors, among others; this is possible because even though their
architectures may vary widely, they all rely on a memory hierarchy to speed up
memory accesses, precisely the hardware feature that \brainslug targets.

\begin{figure}[t]
	\centering
	\includegraphics[width=\linewidth]{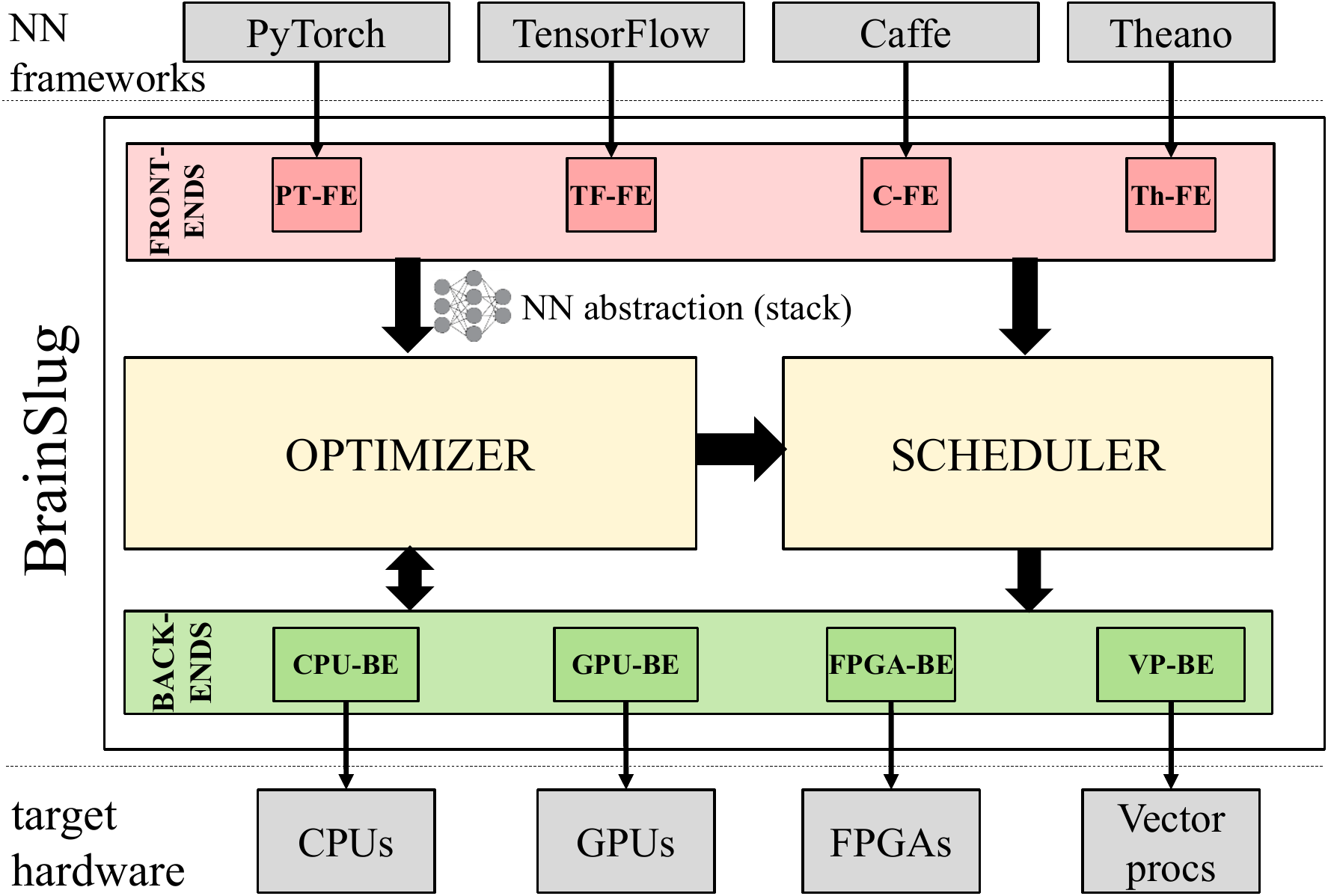}%
	\vspace{1em}%
	\caption{\brainslug architecture consiting of front-ends (FE) that plug to
          existing frameworks and convert their NNs into a common
          abstraction; the optimizer that generates the code to transparently
          accelerate them; and the scheduler that executes such code, relying on
          the back-ends (BE) to run it on different target hardware.}%
	\label{fig:brainslug-arch}%
\end{figure}

To comply with these requirements, the \brainslug architecture introduces the
notion of \emph{front-ends} to support different NN frameworks, and
\emph{back-ends} to be able to execute on different kinds of hardware (see
\autoref{fig:brainslug-arch}).

The \brainslug front-ends are specific to a particular framework. They are in
charge of parsing the NN in whatever format it is in, and providing an
abstraction for it called a \emph{stack} for \brainslug's optimizer component to
use. Further, the front-ends provide glue to invoke \brainslug's scheduler
component whenever the framework launches the prediction process.

The back-ends provide the necessary glue to have \brainslug-generated code
execute on different kinds of hardware, including providing hardware specs to
the optimizer component to help it in generating the code.

Beyond these front and back-ends, \brainslug consists of two main components:
an optimizer, corresponding to a \emph{compile} phase, and a scheduler, mostly
in charge of the \emph{execution} phase. Next, we cover each of these phases in
turn, pointing out how the various components in the \brainslug architecture
interact to optimize and execute a NN. We end the section by giving a more
detailed description of the PyTorch frond-end we implemented along with its API;
and a discussion of the CPU and GPU back-ends.

\subsection{Compile Phase}

\begin{figure}[t]
	\centering
	\includegraphics[width=\linewidth]{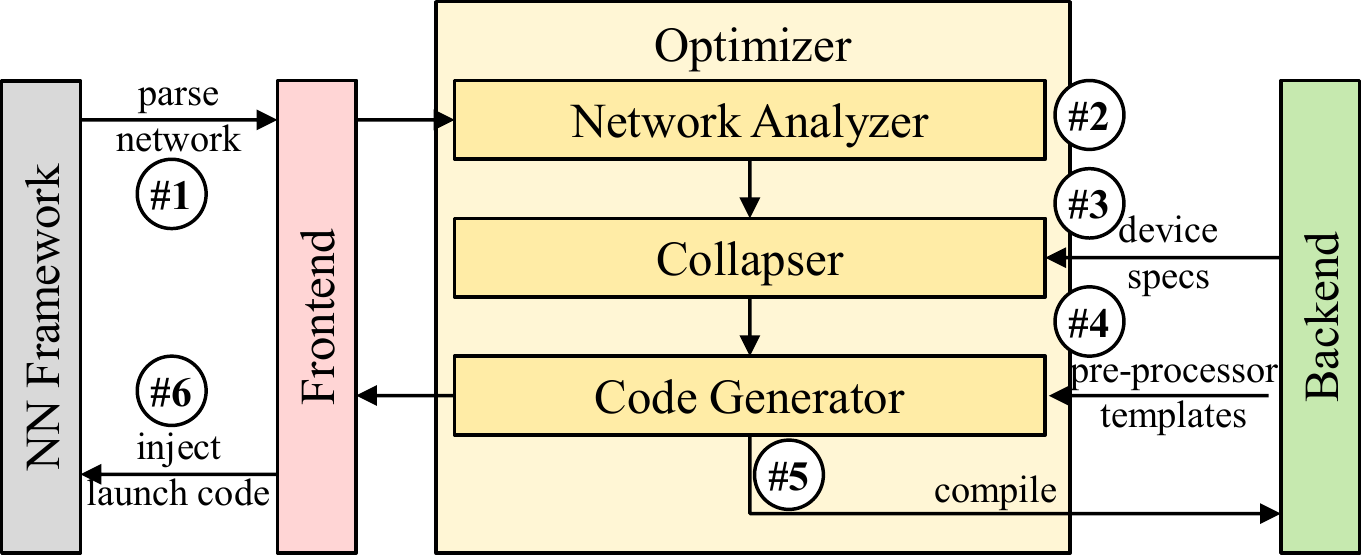}%
	\vspace{1em}%
	\caption{\brainslug's compile phase. The system parses a NN to identify
		  collapsible layers (steps 1, 2), collapses those layers to fit in
		  cache (3), inserts hardware-optimized functions into the code (4),
		  compiles the code (5) and transparently injects the code back into the
		  NN framework (6).}%
	\label{fig:brainslug-compile}%
\end{figure}

\autoref{fig:brainslug-compile} shows \brainslug's compile phase,
which is primarily carried out by \brainslug's optimizer. The process
begins when the NN framework calls the front-end's \texttt{optimize}
function method (Step 1 in the Figure). Next, the Network Analyzer
goes through the neural network and identifies sequences of
optimizable layers; a layer is optimizable if its type is in
\brainslug's list of optimizable layers (Step 2). 

Third, the Collapser retrieves device specs from the back-end(s)
(e.g., cache sizes) and takes care of reducing the layers so that
their memory usage can be fit into a target cache (Step 3). In the
next step (Step 5), the Code Generator retrieves device-specific
pre-processor templates (Step 4) to speed up particular functions (e.g., the
max function maps to \texttt{fmaxf} on a GPU), generates optimized
code and compiles it (Step 5). Finally, the Code Generator uses the
front-end to inject the code back into the NN framework (Step 6).

\begin{figure}[t]
	\centering
	\includegraphics[width=\linewidth]{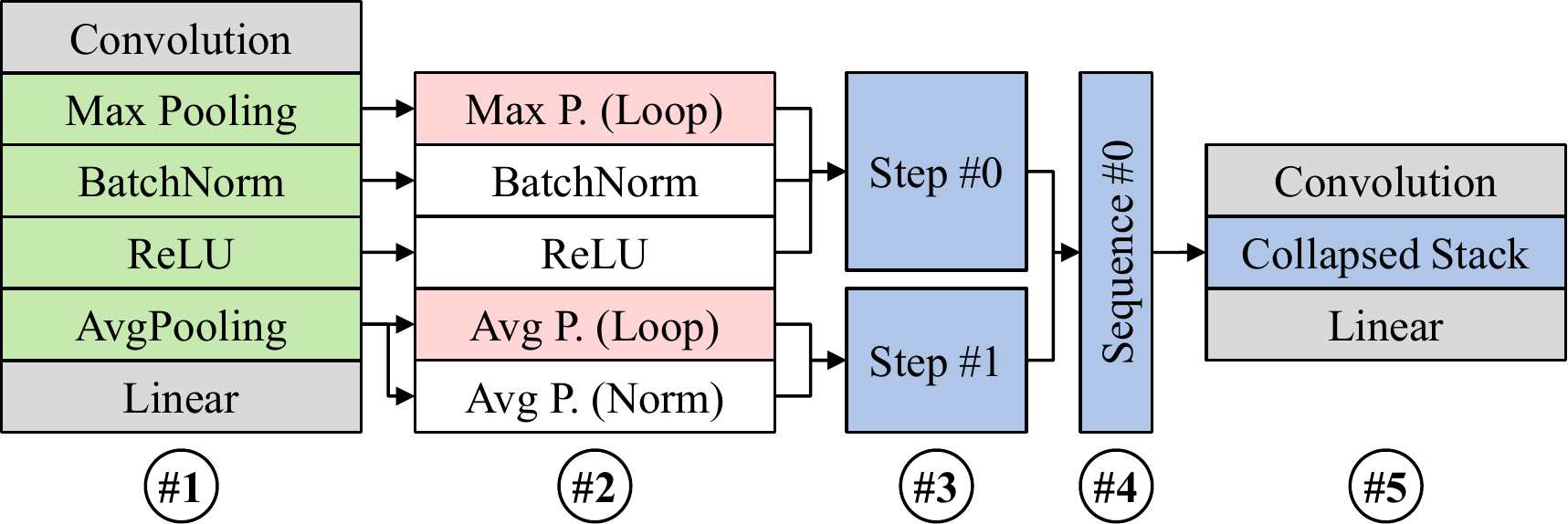}%
	\vspace{1em}%
	\caption{\brainslug's collapse process makes the execution
          of layers, and particularly the data they need, amenable to
          the available cache sizes. Convolution and linear
          layers cannot be optimized and are left
          untouched. Operations marked in red are
          non-element-wise.}%
	\label{fig:brainslug-collapse}%
\end{figure}

\noindent\textbf{Collapse Process.} The collapse process merits
further description (see \autoref{fig:brainslug-collapse} for a diagram
and \autoref{lst:collapse} for corresponding pseudo-code). To begin
with, we identify optimizable layers and group them into a
\emph{stack}. We then map those layers onto basic computational
operations: these can either be element-wise (e.g., Batch
Normalization or ReLU) or non-element-wise (e.g., pooling).

Third, we assign these operations to \emph{steps}. If an operation is
element-wise, it can always be added to a step. If not, it can still
be added to the step if there is not already another non-element-wise
operation present. If this criterion is not met then we create another
step: this is necessary as a non-element-wise layer's operations depend on the output
of a larger number of previous operations.

\begin{lstlisting}[caption={Pseudo code explaining \brainslug's collapse process.}, label={lst:collapse}, float=t]
class Stack:
def collapse():
	#3: group operations in steps
	steps = []
	step  = new Step()
	foreach(operation in optimizable):
		if(not step.onlyElementwise() or
			not operation.isElementwise()):
			step.add(operation)
		else:
			steps.add(step)
			step = new Step()

	#4: group steps in sequences
	sequences = []
	sequence  = new Sequence()
	foreach(step in steps):
		sequence.add(step)
		if(sequence.resourceConsumption() > 
				device.resourceLimit()):
			sequence.remove(step)
			sequences.add(sequence)
			sequence = new Sequence()

#1: identify optimizable layers
stack = new Stack()
foreach(layer in graph):
	if(isOptimizable(layer)):
		#5: replace layers with stack
		if(stack.empty()):
			graph.insertBefore(layer, stack)
		graph.remove(layer)

		#2: map layer to operations
		foreach(operation in layer);
			stack.add(operation)
	else:
		stack.collapse()
\end{lstlisting}
\begin{lstlisting}[caption={Example code of a collapsed stack.}, label={lst:finalcode}, language=C++, morekeywords={foreach,in}, float=t]
void step_0(in_data, out_data, ...):
	foreach(o in out_data):
		foreach(i in in_data):
			MaxPooling()
	BatchNorm()
	ReLU()

void step_1(in_data, out_data, ...):
	foreach(o in out_data):
		foreach(i in in_data):
			AvgPooling()
	AvgNormalization()

void sequence_0(in_data, out_data, ...):
	float cached_data[...]
	step_0(in_data, cached_data, ...)
	step_1(cached_data, out_data, ...)
\end{lstlisting}
\begin{lstlisting}[caption={Snippet showing how to use \brainslug's PyTorch front-end. Only lines 2 and 8 need to be added by the user.}, label={lst:python}, float=t]
import torchvision.models as models
import brainslug

# load the model
model = models.__dict__['...']()

# optimize with BrainSlug
brainslug.optimize(model)

# execute the model
model(...)
\end{lstlisting}

After this, we group the steps in order to utilize hardware resources
efficiently. As each step requires that it is synchronized after it is
processed, all data needs to be stored in a local data cache to pass data from
one step to another. To accomplish this, we bundle these steps into
\emph{sequences}. We iterate over the steps and evaluate if their resource
consumption fits the limitation of the target hardware (e.g., a L1 cache on a
CPU or shared memory on a GPU). The resource consumption is calculated by the
amount of data that each step requires and the number of active SIMD units of
the processor that share this data. For example: If we have 128 SIMD units, a
non-overlapping pooling layer with kernel size 3x3, and 32 channels, we would
require 128*32\,B for the output and 128*3*3*32\,B for the input. An additional
layer would have the previous input size as output, and a corresponding larger
input size. If the stacked steps do not exceed the hardware resources, we add
the step to the sequence, otherwise we create a new sequence.

Finally, we generate the final code. There are two possible scenarios. First, if
a sequence only contains a single step, we iterate over the entire input data:
in this case data locality is achieved by directly passing the values from one
operation to another. If there are multiple steps, we need to perform the
previously-mentioned synchronization between the steps. In the case that the
cache size limit is not reached, we increase the size of it, so that each SIMD
unit may not calculate a single output value, but multiple ones, to better
utilize the given hardware resources. Finally, we compile the code using a
device-specific compiler and replace the optimized layers in the NN with our
collapsed stack. To illustrate, \autoref{lst:finalcode} shows how the example in
\autoref{fig:brainslug-collapse} is mapped onto the actual final code.

\subsection{Execution Phase}
\brainslug's execution phase, embodied by the scheduler, handles the execution
of the compute kernels. When a stack is executed, the front-end gathers all
necessary data and parameter tensors. The scheduler then calculates the output
size and allocates memory using the NN framework. After this the kernel function
object (cubin for GPU and dll for CPU) is loaded, executed and the result buffer
is returned to the NN framework. If there is more than one sequence in a stack
the sequences are executed in a serialized fashion.

\subsection{PyTorch Front-end and API} 
We implemented a PyTorch front-end. We chose PyTorch as it was the
first NN framework to support dynamic network graphs that can be reshaped at
runtime. This feature complicates the implementation but allows us to show that
our method can be applied even in such a highly dynamic scenario.

The frontend parses through the neural network, groups all optimizable layers in
stacks and passes these to the \brainslug optimizer. These are then removed from
the network and replaced by a special \brainslug layer (one per stack) that pass
the control flow to the \brainslug scheduler whenever they are triggered. If
there are multiple equivalent stacks, \brainslug only generates the code once
and reuses it for all identical stacks.

Finally, it is worth noting that the front-end is extremely easy to
use: the user need only add a few lines of code in order to enable
transparent acceleration of the neural networks (see \autoref{lst:python}).

\subsection{CPU and GPU Backends}
\noindent\textbf{GPU}: GPUs are often used for processing of neural networks,
mainly because of their high compute performance and memory throughput. For the
implementation we use two code building blocks.

For steps that only perform element-wise operations, we start as many thread
blocks as there are channels and each thread blocks applies its calculations on
each batch, for a specific channel.

For pooling layers we distinguish between stacked and non-stacked. In the
non-stacked case, we start \texttt{BatchSize * Channels} thread blocks. The SIMD
units iterate over all data elements, while we process as many element as we
have SIMD units in parallel. In the stacked case, we use \texttt{BatchSize *
Channels * Patches} thread blocks, where each patch represents a depth-first
processing and use the SIMD units the same way. To store the data for the
depth-first processing we use two buffers allocated in the devices shared
memory. When a step is processed, we synchronize the entire thread block and
swap the buffers for the next step.

In general we use a thread count of 128, which is a good trade-off between
overhead when synchronizing a thread block and compute utilization. Further, we
limit the usage of shared memory to 16\,kB (depending on the GPU either 64 or
92\,kB would be available), as this can have a negative impact on the
performance because it reduces the amount of blocks that can be scheduled onto the
GPU multiprocessor, resulting in less opportunities to employ latency hiding.

\noindent\textbf{CPU}: The CPU back-end relies on the \textit{Intel SPMD Program
Compiler}~\cite{intelispc} (ISPC). ISPC can be seen as
``CUDA for CPUs'': it adds syntactic sugar to the application and explicitly
defines which computations should be done by a single SIMD unit. Because of the
similarities between ISPC and CUDA we can share many parts of the implementation
between both architectures. As CPUs do not have a dedicated shared memory, we
allocate the memory on the stack. The other parts of the implementation are
 similar and differ only in the way variables are used -- either by all or only single SIMD units -- and replacing the outer loops with the
ISPC specific \texttt{foreach(...)} instruction. ISPC can target different
instruction sets, e.g. SSE[2,4], AVX[1,2] and AVX512 for Intel's Knight's
Landing. We use the default values from the compiler. ISPC further supports a task
system, similar to CUDA thread blocks. This task system has to use some predefined
variants provided by ISPC or can be implemented by the user. As our task system
does not require any special features, we implemented a simple variant
based on \texttt{parallel for} using Intel's Threading Building
Blocks~\cite{inteltbb} to minimize the framework's
overhead.

%% file: eval.tex
\section{Evaluation}
\label{sec:eval}
To evaluate \brainslug we chose the
TorchVision\cite{torchvision} package.
TorchVision contains a series of broadly used neural network architectures for
computer vision applications. We use the entire set of available networks
ranging from AlexNet (A)~\cite{alexnet}, Densenet-(121, 161, 169, and 201) (D)
\cite{densenet}, Inception v3 (I)~\cite{inception}, Resnet-(18, 34, 50, 101, and
152) (R)~\cite{resnet}, Squeezenet-(1.0 and 1.1) (S)~\cite{squeezenet}, and
VGG-(11, 13, 16, and 19, with and without Batch Normalization) (V)~\cite{vgg}, a
total of 21 different architecture and parameter combinations.

We run all tests on a server with an Intel Xeon E5-2690v4, an NVIDIA GeForce GTX
1080 Ti, Debian 9, NVIDIA GPU driver v384.81, CUDA v9.0, ISPC v1.9.2, Python
v3.5.3 and PyTorch v0.3.0 (using cuDNN). We perform the test times ten times for
the GPU and five times for the CPU and we take the minimum execution time for
both PyTorch and \brainslug results.

\begin{figure}[t]
	\centering
	\includegraphics[clip, trim=0 0 0 0, width=\linewidth]{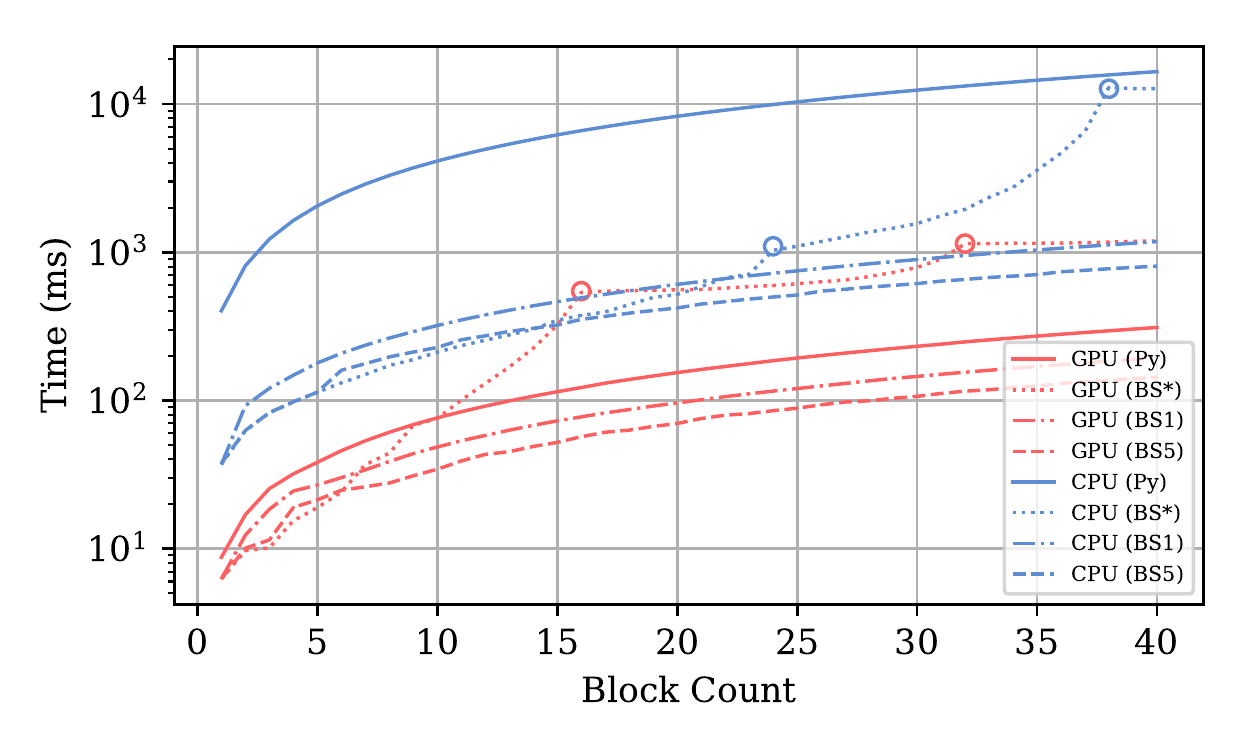}
	\caption{Performance of \brainslug's stacked layer mechanism versus PyTorch
		for increasing numbers of $<$Max-Pooling,Batch Normalization,ReLU$>$
		blocks.}
	\label{fig:eval:synthetic_scaling}
\end{figure}

\subsection{Stacked Layers Acceleration}
For the first experiment we want to evaluate the advantage of our proposed layer
stacking mechanism. To do so, we build artificial neural networks consisting only of
layers that can be optimized using \brainslug. In particular, we define a block
consisting of a Max-Pooling (Kernel: 3x3, Stride: 1x1 and Padding: 1x1), a Batch
Normalization and a ReLU layer, and create neural networks that comprise between
1 and 40 of these blocks. We execute these networks on a CPU and GPU (see
\autoref{fig:eval:synthetic_scaling}, notice the log scale) and evaluated three
different strategies: only 1 step per sequence, max 5 steps per sequence and
unrestricted. 

On the CPU, the PyTorch implementation is always 10-20x slower than
\brainslug. This massive increase is partially due to the fact that
the current PyTorch implementation is not particularly optimized for
CPUs: most significantly, it does not use any explicit vector
processing instructions. In contrast, we use ISPC for vector
operations, so in theory we have 8x more computational power (AVX2
with 8x 32Bit float operations). Further, the PyTorch CPU code relies
on OpenMP \texttt{parallel for} constructs, but does not define a
specific execution schedule, yielding sub-optimal
performance. On the GPU, \brainslug yields a speed-up of 1.4-2.2x.

\begin{figure*}[t]
	\begin{minipage}[t]{0.49\linewidth}%
	\includegraphics[clip, trim=0 0 0 0, width=\linewidth]{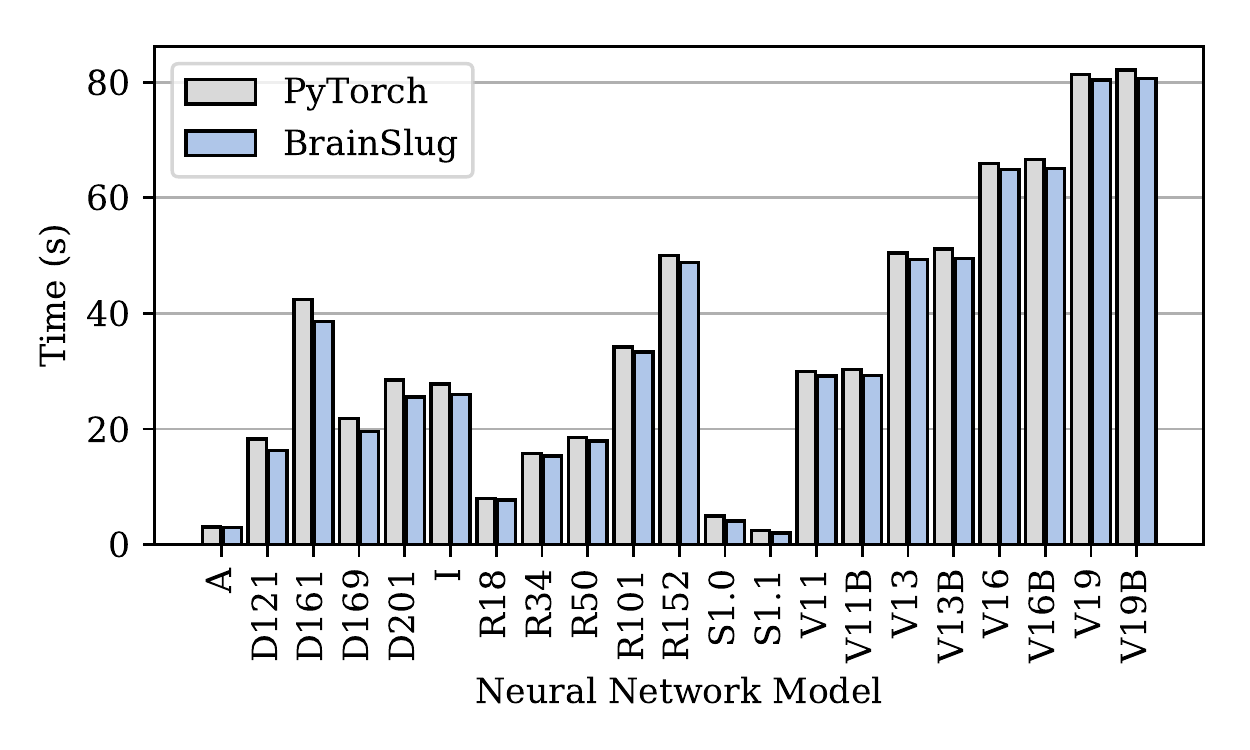}%
	\caption{\label{fig:eval_total_cpu_time}Comparison of calculation time between PyTorch and \brainslug for TorchVision's neural networks (CPU, batch size 128).}%
	\end{minipage}%
	\hspace{0.02\linewidth}%
	\begin{minipage}[t]{0.49\linewidth}%
	\centering
	\includegraphics[clip, trim=0 0 0 0, width=\linewidth]{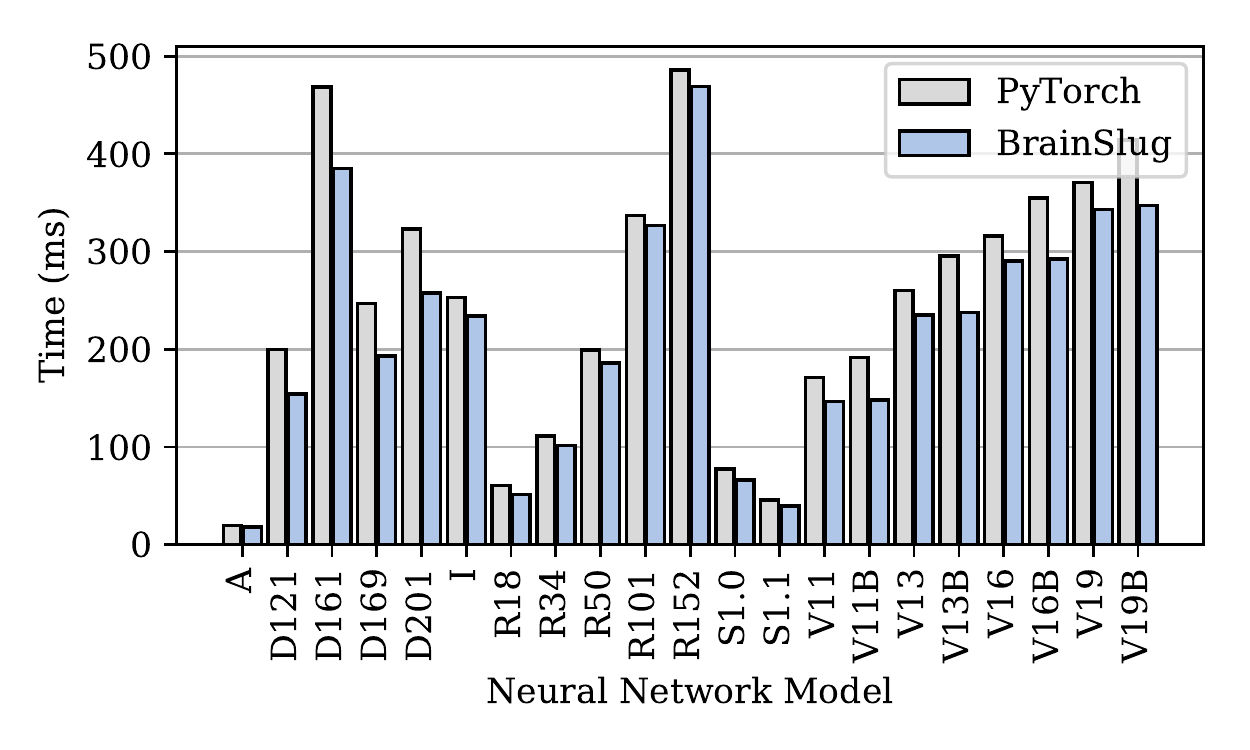}%
	\caption{\label{fig:eval_total_gpu_time}Comparison of calculation time between PyTorch and \brainslug for TorchVision's neural networks (GPU, batch size 128).}%
	\end{minipage}%

	\begin{minipage}[t]{0.49\linewidth}%
	\includegraphics[clip, trim=0 0 0 0, width=\linewidth]{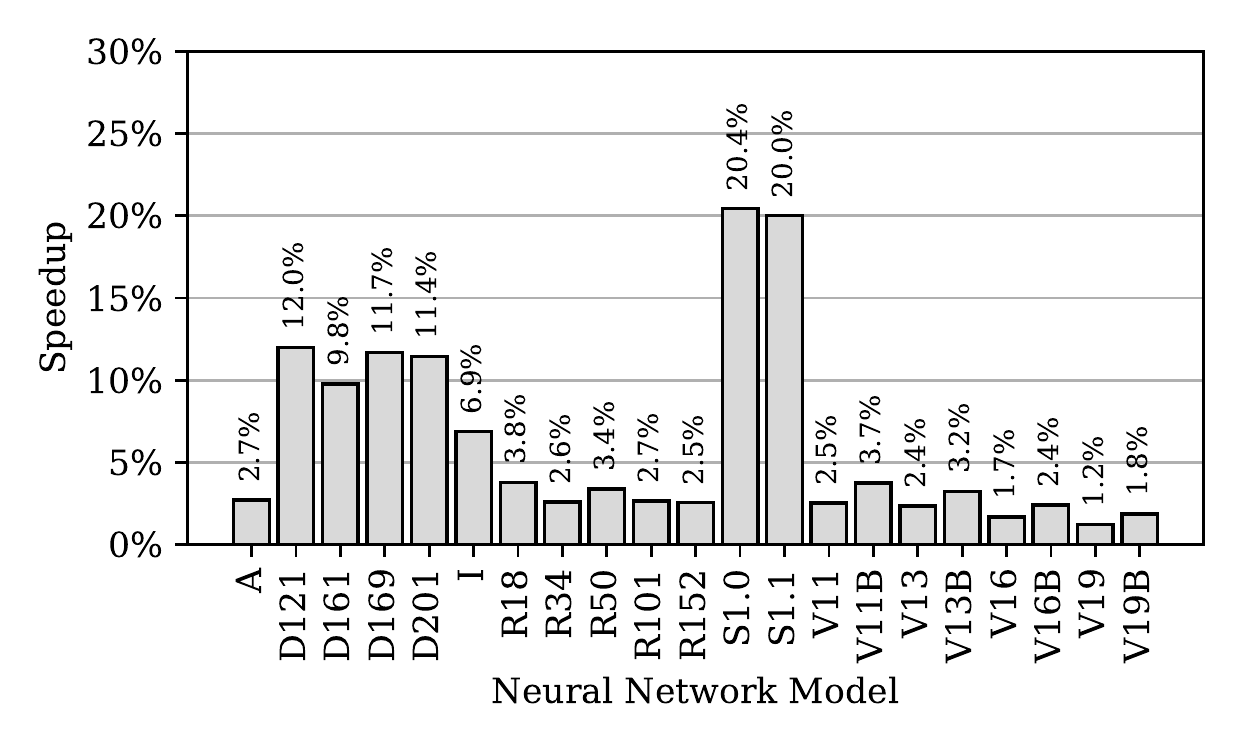}%
	\caption{\label{fig:eval_total_cpu_speedup}\brainslug's speed-up over PyTorch of TorchVision neural networks (CPU, batch size 128).}%
	\end{minipage}%
	\hspace{0.02\linewidth}%
	\begin{minipage}[t]{0.49\linewidth}%
	\centering
	\includegraphics[clip, trim=0 0 0 0, width=\linewidth]{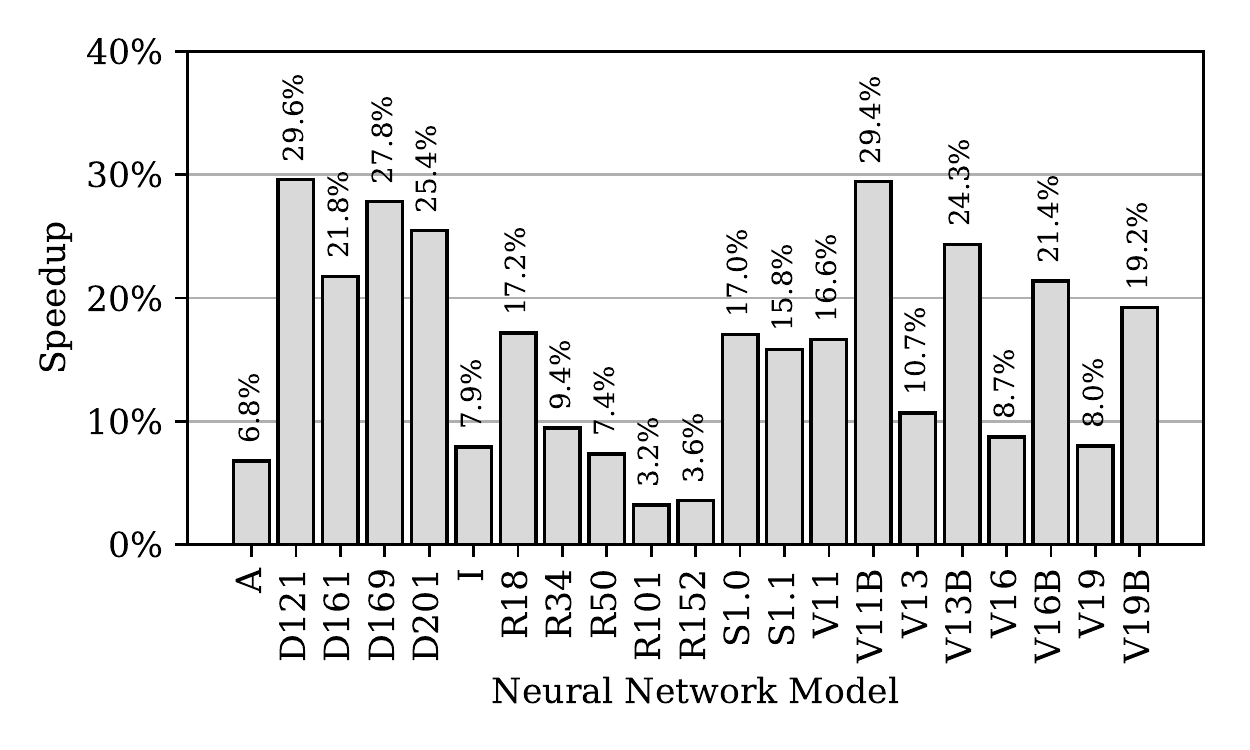}%
	\caption{\label{fig:eval_total_gpu_speedup}\brainslug's speed-up over PyTorch of TorchVision neural networks (GPU, batch size 128).}%
	\end{minipage}%
\end{figure*}

For both devices the performance improves even if we only allow one step per
sequence. It further improves when we stack multiple steps in a sequence, up to
61\% for the GPU and 58\% for the CPU. In the unrestricted case, we can see that
for the lower block counts it is equal or even slightly better than the 5 step
scenario. However, the performance significantly decreases for larger values
until it reaches an artifact (indicated by circles) that happens for the GPU at
16 and 32, and for the CPU at 24 and 38. These artifacts occur whenever the
cache size limit is reached and an additional sequence is required. The cause
for this increase in required cache size is the padding value of the Max-Pooling
layer. This causes an overlap of data and, as previously discussed for the
convolutional layers, results in redundant operations. As each block adds new
padding, the value increases with each additional block. The performance
improves at these points since the new sequence does not suffer from the
redundant operations in the first place, but only if too many blocks are added
to it.

\subsection{Full Network Acceleration}
Next, we evaluate the acceleration that \brainslug provides when
executing more realistic neural
networks. Figures~\ref{fig:eval_total_cpu_time}
and~\ref{fig:eval_total_gpu_time} show the total execution time when
running the networks with a batch size of 128 for CPUs and GPUs
respectively, while Figures~\ref{fig:eval_total_cpu_speedup} and
\ref{fig:eval_total_gpu_speedup} show the relative speedup with
respect to PyTorch.

While the networks have significantly varying execution times ranging
from very short (AlexNet) to quite long (Densenet-161 and Resnet-152),
\brainslug provides a speed up in all cases, with the most pronounced
improvements for Densenets on both CPU and GPU, VGGs with Batch
Normalization on GPU, and Squeezenets on CPU.  Note that adding the
Batch Normalization layer to the VGG networks has a significant impact
on PyTorch's computation time, while there is virtually no change in
\brainslug's case: an effect directly attributable to \brainslug
collapsing the normalization into the previous step.

\autoref{tbl:eval_total_table} shows \brainslug's full speed-up
results for all networks on CPU and GPU for batch sizes from 1 to
256. 
The results clearly indicate that \brainslug outperforms PyTorch on the GPU with
batch sizes bigger than 8 (except for Resnet-101 and \mbox{-152}), and
for all cases for the CPU.

\begin{lstlisting}[caption={PyTorch's Max-Pooling implementation.}, label=lst:maxerror, float=t]
#omp parallel for
foreach(batch):
	#omp parallel for
	foreach(channel):
		...
\end{lstlisting}

The results show large performance gains for small batch sizes when
running on the CPU. This is related to a programming error in
PyTorch's Max-Pooling implementation (see \autoref{lst:maxerror}). The
code uses two nested OpenMP \texttt{parallel for} loops, which means
that only the outer loop is parallelized over the CPU cores. In the
extreme case of batch size $= 1$, the entire function utilizes only a
single core. In \brainslug we use only one \texttt{parallel for} loop,
iterating over \texttt{BatchSize $\times$ Channels} elements, so we
can always leverage parallelism.

Finally, note that negative values for the GPU batch sizes 1–-4 look
significant but in absolute terms they are not: in these cases the
execution time is only a few milliseconds, while for larger ones, it
is hundreds of milliseconds. This relatively performance difference is
mainly because our implementation is optimized towards larger batch
sizes.

\begin{table*}[t]
	\centering
	\includegraphics[clip, trim=0 0 0 0, width=\linewidth]{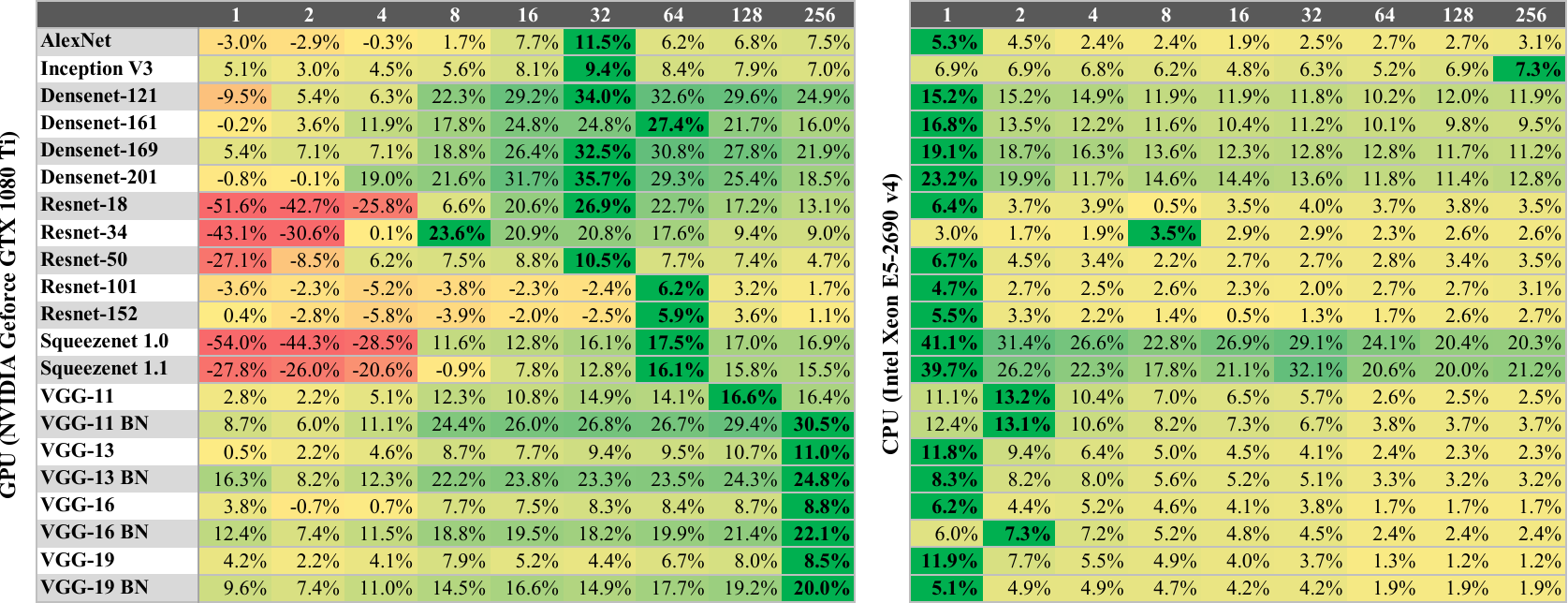}
	\caption{\brainslug full speed-up results compared to PyTorch on a CPU and
		GPU for all neural networks.}
	\label{tbl:eval_total_table}
	\vspace{-3pt}
\end{table*}

\subsection{Detailed Performance Analysis}
\brainslug's performance gains stem from optimizing some layer types,
while leaving others as they are. Here we provide a more detailed
analysis to answer the following question: for real-world neural
networks, how much does \brainslug improve the performance of the
optimizable layers, and what fraction of the overall runtime is this
optimizable part?

\begin{table}[t]
	\centering%
	\input{DetailKernel.tex}	\vspace{1.0em}
	\caption{For each NN at batch size 128: the number of layers, how
          many \brainslug can optimize and into how many stacks.
          Opt.\ Speed Up is the acceleration for the optimizable
          layers, \% of Total Time is the time the optimized layers
          take within the full execution, and Total Speed Up is the
          speed-up for the entire network.}%
	\label{tbl:eval_detailkernel}%
\end{table}

As shown in \autoref{tbl:eval_detailkernel}, the complexity of the networks
ranges from 27 up to 709 layers, with \brainslug able to optimize 44 to 64\% of
them using 8 to 204 stacks. For those layers, we achieve speed-ups of 321.2 to
842.9\% on the CPU and 5.7 to 222.9\% on the GPU (all results are for a batch
size of 128). Again, the speed-up for the CPU is significantly higher than for
the GPU due to PyTorch's less-than-optimal CPU implementation. Overall, the
optimizable layers represent 2.5 to 16.9\% for the CPU and 13.7 to 47.4\% for
the GPU of the total computation time (\% of Total Time columns), with the rest
of the time being spent mostly on convolutional layers. This leads to a total
speed-up of 2.1\% to 13.9\% for the CPU and 1.1\% and 20.9\%.
Note that this speed-up only concerns the pure compute kernel time, and is
hence different from the numbers for batch size 128 in Table~\ref{tbl:eval_total_table};
the total improvement is in fact often higher because, for example, \brainslug needs
fewer memory allocations.

\begin{figure}[t]
	\vspace{-9pt}
	\centering
	\includegraphics[clip, trim=0 0 0 0, width=\linewidth]{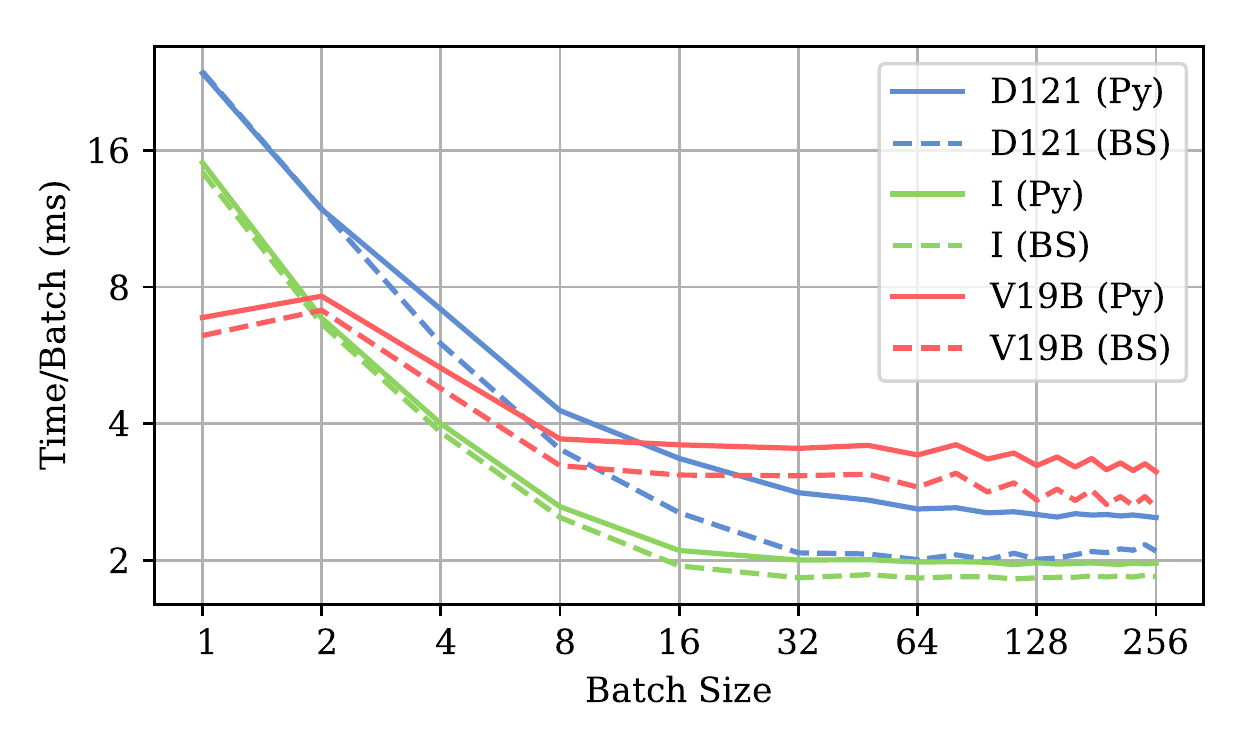}%
	\caption{\label{fig:eval:batch_scaling}Scaling behavior of \brainslug (BS) versus PyTorch (Py) for different batch sizes.}%
\end{figure}

\subsection{Batch Size Scaling Behavior}
One important parameter for neural network performance is the batch size, which
represents the number of independent data parts (e.g., images) that are
processed at the same time. Most operations can operate independently on
individual batches, which can be leveraged for parallelism.
\autoref{fig:eval:batch_scaling} shows how PyTorch and \brainslug
scale with respect to increasing batch sizes for three selected
networks. As can be seen, both scale with batch size but \brainslug
performs always best, showing increasing gains for larger batch sizes.


%% file: DetailKernel.tex
\scriptsize
\setlength{\tabcolsep}{1pt}
\begin{tabular}{l|ccc|S[table-format=3.1]S[table-format=3.1]|S[table-format=2.1]S[table-format=2.1]|S[table-format=2.1]S[table-format=2.1]}
\rowcolor{gray} & \multicolumn{3}{c|}{\textcolor{white}{Network}} & \multicolumn{2}{c|}{\parbox{1.4cm}{\centering \textcolor{white}{Opt.\ Speed-up [\%]}}} & \multicolumn{2}{c|}{\parbox{1.35cm}{\centering \textcolor{white}{\% of Total Time}}} & \multicolumn{2}{c}{\parbox{1.36cm}{\centering \textcolor{white}{Total Speed-up [\%]}}} \\
\rowcolor{gray} \textcolor{white}{Name} & \textcolor{white}{Layers} & \textcolor{white}{Opt.} & \textcolor{white}{Stacks} & \textcolor{white}{CPU} & \textcolor{white}{GPU} & \textcolor{white}{CPU} & \textcolor{white}{GPU} & \textcolor{white}{CPU} & \textcolor{white}{GPU} \\

\rowcolor{white} Alexnet & 27 & 12 & 8 & 483.1 & 72.7 & 3.0 & 13.7 & 2.5 & 5.8 \\
\rowcolor{lightgray} Inception V3 & 316 & 203 & 103 & 451.2 & 27.5 & 8.7 & 28.6 & 7.1 & 6.2 \\
\rowcolor{white} Densenet-121 & 429 & 247 & 124 & 411.1 & 77.3 & 13.1 & 47.3 & 10.5 & 20.6 \\
\rowcolor{lightgray} Densenet-161 & 569 & 327 & 164 & 446.9 & 68.3 & 10.9 & 38.6 & 8.9 & 15.7 \\
\rowcolor{white} Densenet-169 & 597 & 343 & 172 & 414.0 & 72.6 & 13.9 & 47.4 & 11.2 & 19.9 \\
\rowcolor{lightgray} Densenet-201 & 709 & 407 & 204 & 415.1 & 66.5 & 13.7 & 47.1 & 11.0 & 18.8 \\
\rowcolor{white} Resnet-18 & 71 & 39 & 21 & 387.5 & 76.6 & 4.8 & 22.2 & 3.8 & 9.6 \\
\rowcolor{lightgray} Resnet-34 & 127 & 71 & 37 & 436.4 & 60.0 & 3.6 & 17.9 & 2.9 & 6.7 \\
\rowcolor{white} Resnet-50 & 177 & 104 & 54 & 348.5 & 13.7 & 8.2 & 24.6 & 6.4 & 3.0 \\
\rowcolor{lightgray} Resnet-101 & 347 & 206 & 105 & 321.2 & 8.2 & 6.5 & 20.7 & 5.0 & 1.6 \\
\rowcolor{white} Resnet-152 & 517 & 308 & 156 & 319.2 & 5.7 & 6.3 & 19.5 & 4.8 & 1.1 \\
\rowcolor{lightgray} Squeezenet 1.0 & 66 & 31 & 29 & 842.9 & 34.7 & 4.5 & 29.3 & 4.0 & 7.5 \\
\rowcolor{white} Squeezenet 1.1 & 66 & 31 & 29 & 457.1 & 32.3 & 16.9 & 33.5 & 13.9 & 8.2 \\
\rowcolor{lightgray} VGG11 & 35 & 17 & 10 & 808.3 & 113.5 & 3.8 & 21.5 & 3.3 & 11.4 \\
\rowcolor{white} VGG11 BN & 43 & 25 & 10 & 842.9 & 222.9 & 4.5 & 30.2 & 4.0 & 20.9 \\
\rowcolor{lightgray} VGG13 & 39 & 19 & 12 & 661.9 & 59.7 & 3.3 & 19.4 & 2.8 &  7.3 \\
\rowcolor{white} VGG13 BN & 49 & 29 & 12 & 622.2 & 159.2 & 4.0 & 29.5 & 3.4 & 18.1 \\
\rowcolor{lightgray} VGG16 & 45 & 22 & 15 & 665.2 & 51.7 & 2.7 & 17.2 & 2.4 & 5.9 \\
\rowcolor{white} VGG16 BN & 58 & 35 & 15 & 620.0 & 146.7 & 3.4 & 26.5 & 2.9 & 15.8 \\
\rowcolor{lightgray} VGG19 & 51 & 25 & 18 & 650.0 & 44.8 & 2.5 & 15.6 & 2.1 & 4.8 \\
\rowcolor{white} VGG19 BN & 67 & 41 & 18 & 618.2 & 137.7 & 3.0 & 24.6 & 2.6 & 14.2 \\

\end{tabular}

%% file: related.tex
\section{Related Work}
\label{sec:related}
In this work we focus on accelerating the forward pass in deep
neural networks on both CPUs and GPUs. Due to the frequent occurrence
of convolutional and dense layers in these networks (see
Figure~\ref{fig-cnn}), recent work has focused specifically on
improving the multiply-and-accumulate (MAC) operations prevalent in
these layers. 
CPUs and GPUs have libraries that support SIMD or SIMT-based processing such as
Intel's ``single program, multiple data'' (SPMD) compilers. In general, the MAC
operations resulting from convolutional and dense layers can be mapped to
multiplications between two matrices. Libraries such as
cuDNN~\cite{chetlur2014cudnn} and cuBLAS~\cite{cublas} for GPUs and Intel
MKL~\cite{intelmkl} and OpenBLAS~\cite{openblas} for CPUs are optimized for
matrix--matrix operations. Moreover, there are specialized algorithms that can
lead to speed-ups for the multiply operation in convolutional layers. For
instance, performing a fast Fourier transform (FFT) has been shown to be
beneficial for convolutional layers with certain properties~\cite{mathieu:2014}.
There are also several other approaches that reduce the number of expensive
operations required for matrix-matrix multiplications. Examples are the
application of Strassen's~\cite{cong2014minimizing} and Winograd's
algorithms~\cite{lavin2016fast} for accelerating the processing of convolutional
layers. Deep learning libraries such as NVIDIA's cuDNN and
TensorRT~\cite{tensorrt} utilize heuristics for choosing the algorithmic method
expected to work best for a given convolutional layer. TensorRT selects
different implementations according to the used hardware and optimizes memory
allocation. Due to the extensive engineering that goes into the design of these
heuristics and implementations, the processing of convolutional layers is highly
optimized and hard to improve on; consequently, \brainslug focuses on
improvements to other commonly-used layer types.

The main disadvantage of TensorRT is that it only works if all used layers are
known to the framework, as it directly translates the entire network
into its own implementation for NVIDIA GPUs. \brainslug, in contrast, only replaces
parts of the network that it knows. This enables us to create and explore
user-created layers and still benefit from \brainslug's
improvements. Further, \brainslug is designed as an extensible platform,
allowing users to apply its optimizations not only to GPUs but to all kinds of
processors and accelerators.

The algorithmic optimizations discussed so far do not change the network
architecture or the result of the computations. There are several algorithmic
tricks one can apply to trade-off accuracy for efficiency. For instance,
TensorRT can reduce the precision from 32\,Bit floating point to 16-Bit floating
point or even 8-Bit integer, which improves performance but might decrease
accuracy. It is even possible to work with binarized neural networks, that is,
networks that perform binary instead of floating point
operations~\cite{hubara2016binarized,courbariaux2016binarized,rastegari2016xnor}.
There are numerous other methods that change the structure and parameters of the
original DNN to improve performance. For instance, it is possible to prune
filters during the learning process which reduces the amount of computation
required for the convolutional layers~\cite{li2016pruning}. In a different line
of work, the network units with low-valued activations are pruned. This was
shown to result in a 11\% speed up~\cite{albericio2016cnvlutin} or a
substantial reduction in power consumption~\cite{reagen2016minerva}. 
With \brainslug, we focus on
optimizations that do not alter the original DNN: both the original
and optimized DNN perform exactly the same operations on the hardware.

Alwani et al.~\cite{2016:alwani:fused} proposed to fuse layers of
convolutional neural networks for faster processing on FPGAs, merging
the first two convolutional layers of a neural network. Their
method uses a data shifting approach to reduce the recomputation of
overlapping data regions. This is quite efficient for FPGAs but is difficult to implement on CPUs and GPUs, and has the limitation of only being applicable to no more than two convolutional layers. As already mentioned, our method
does not focus on convolutional layers but accelerates the entire
network by aggregating consecutive non-convolutional layers.

Sze et al.~\cite{sze2017efficient} discuss different methods for
energy-efficient dataflows on neural network accelerators, suggesting
the development of a specialized neural network accelerator with a
mesh-based processing architecture. In contrast, our method targets
off-the-shelf, cheap hardware that provides excellent compute power per
dollar.

%% file: discussion.tex
\section{Discussion and Future Work}
\label{sec:discussion}

We have shown that \brainslug (\href{http://brainslug.info}{brainslug.info}) accelerates commonly used deep neural networks by
as much as 41.1\% on CPUs and 35.7\% on GPUs while requiring
minimal code changes. These improvements are significant considering that training
such networks on big data can take up to several weeks. {\brainslug}'s speed-up is most  
pronounced for the more commonly used batch sizes of 8 and up. For instance, the \textsc{DenseNet} architectures are usually trained with a per-GPU batch size of 32~\cite{densenet}, a batch size where \brainslug achieves the best performance improvement. Due to recent results insights into the benefits of increasing the batch size during training~\cite{smith2017don} and the generally growing size of main memory on GPUs, we expect training batch sizes to further increase in the future. 

\noindent\textbf{Extending \brainslug:} \brainslug is designed to make
it easy to extend and, as such, provides APIs that need to be
implemented when developing front and back-ends. Adding a new
front-end requires the biggest effort, since it has to parse through
the NN graph and identify optimizable layers; this cannot be
implemented generically as every NN framework uses a different
representation. Further, it is necessary to connect \brainslug's
runtime system with the NN framework, so that \brainslug can interact
with framework-specific data structures. For PyTorch, our front-end
implementation consists of 270 lines of Python code and 438 lines of
C++. To add a new back-end requires much less work, as only methods to
load and execute the device code are required. In our implementation,
we have 299 lines of C++ code (code + header files) for NVIDIA GPUs
(including integration of the NVIDIA profiling library) and 165 lines of code for CPUs.

\noindent\textbf{Limitations:} Although in theory our stacking method
can be applied to several different kinds of layers, we figured out
that in certain cases it is not beneficial. While we were able to
achieve significant speed ups for element-wise and pooling layers, we
have not been able to improve linear and convolutional layers. For
convolutional layers the problem is that the operation itself uses
overlapping data areas. Because of this overlap, \brainslug would
force neighboring data paths to have to do redundant calculations. As
convolution is already a compute-bound operation, and since \brainslug
optimizes memory accesses and not the actual computation, these
redundant calculations reduce overall performance. For linear layers
the problem is more conceptual. A linear layer can be represented by a
matrix--matrix multiplication. Instead of executing this as a
matrix--matrix multiplication, \brainslug would strip it down to
multiple vector--matrix multiplications. The problem of this method is
that the processor needs to load the entire weight matrix for each
output vector. In contrast, in a matrix--matrix multiplication the
weight matrix can be significantly better reused, resulting in less
memory transactions compared to multiple vector--matrix multiplications.

\noindent\textbf{Future Work:} We plan to enhance \brainslug by
adding more front-ends to support a larger variety of frameworks. Further, we
plan to expand our optimizations to training, as this is the most time consuming
operation for neural networks; we expect \brainslug to be able to achieve
equivalent speed-ups for it. Finally, we are also targeting additional types of
hardware including vector or neural network processors.

%% file: references.tex
\bibliographystyle{acm}
\bibliography{main}